\documentclass[amstex,aps,preprint,eqsecnum]{revtex4}

\usepackage{amsmath}
\usepackage{graphics}

\begin{document}

\title{The Expansion of the Universe and the Cosmological Constant
Problem}
\author{R. F. O'Connell}
\affiliation{Department of Physics and Astronomy, Louisiana State University, Baton
Rouge, LA 70803-4001}

\begin{abstract}
The discovery that the expansion of the universe is accelerating
in time is a major discovery which still awaits adequate
explanation.  It is generally agreed that this implies a cosmic repulsion as a
result of the existence of a cosmological constant $\wedge>0$.  However, estimates
of $\wedge$, based on calculations of the zero-point fluctuations of quantum
fields are too large by over a hundred orders of magnitude.  This
result is obtained by summing the zero-point energies up to a large cutoff energy
$\Omega$, based on the Planck scale.  Since there is no compelling reason for this
choice, we argue that since all known quantum electrodynamic (QED) effects involves
interaction with matter, a preferred choice should be based on causality
and other considerations, leading to a much lower value for $\wedge$.
\end{abstract}

\date{\today}

\maketitle

Measurements of the expansion of the universe \cite{tegmark} imply a vacuum energy
density which is over a hundred orders of magnitude too small compared to current
theoretical calculations \cite{wein, peebles} which attribute this so-called dark
energy to quantum zero-point energies.  The latter embrace a variety of fields but
the essence of our remarks here can be captured by concentrating on just the normal
modes of the electromagnetic field.  The vacuum energy density for the E-M fields is
simply given by summing the zero-point energies $(\hbar\omega/2)$ of all the normal
modes, up to a cutoff frequency $\Omega$, to get (after inclusion of a factor 2
arising from the fact that there are two normal modes of the E-M field for each wave
vector)

\begin{equation}
<\rho>=2\int^{p_{max}}_{0}~dp~\frac{4\pi{p^{2}}}{(2\pi\hbar)^{3}}~~\frac{\hbar\omega}{2},
\end{equation}  where $p_{max}$ is the momentum cutoff.  Hence, since $p=(\hbar
\omega/c)$, we obtain,

\begin{equation}
<\rho>=\frac{\hbar\Omega^{4}}{8\pi^{2}c^{3}}.
\end{equation}  The current wisdom is
to argue that, if general relativity is valid up to the Planck scale, then one
might guess that $\Omega$ is given by $\Omega_{p}$ where

\begin{equation}
\hbar\Omega_{p}=E_{p}=\left(\frac{\hbar{c}^{5}}{G}\right)^{1/2},
\end{equation}  leading to a vacuum energy density, $<\rho_{p}>$ say, which is
over a hundred orders of magnitude too large \cite{wein, peebles}.

Here, we argue that allowing $\Omega$ to have such a large value is completely ad
hoc and not based on compelling physical arguments.  Instead, we take the point of
view that one should not consider vacuum fluctuations in isolation but rather in
interaction with matter fields and, as a consequence, their contribution to the
energy of the vacuum is much less.  The fact is that all the well-known observed
QED effects (Lamb shift, Casimer effect, etc.) involve interation of the vacuum
field with matter.  With that in mind, we prefer to use atomic units ($\hbar=M=e=1$
\cite{bethe}, where $M$ and $e$ refer to the mass and charge of the electron,
respectively).  In these units, consistent with the fact that the atomic unit of
velocity is
$2.1877\times{10^{8}}cm/s$, we note that
$c=\alpha^{-1}=137.036$, where $\alpha$ is the fine-structure constant.  Also, since
$(GM^{2}/e^{2})=2.401\times{10}^{-43}$, we see that in these units
$G=2.401\times{10}^{-43}=1.309~\alpha^{20}$.   We note that $\alpha$ is the
natural expansion parameter in QED.  Thus, in atomic units (a.u.), (2) and  (3)
simply become 

\begin{equation}
<\rho>=\frac{\alpha^{3}}{8\pi^{2}}~\Omega^{4}=4.92\times{10^{-9}}~\Omega^{4}
\end{equation} and

\begin{equation}
\Omega_{p}=\alpha^{-5/2}~G^{-1/2}(a.u.)  = 
0.874~\alpha^{-25/2}=4.49\times{10}^{26}.
\end{equation}  Thus, from (4) and (5), we obtain

\begin{equation}
<\rho_{P}>=2\times{10^{98}}a.u.=1.2\times{10}^{96}g/cm^{3}.
\end{equation}  This is to be compared with the measured vacuum energy density 
\cite{tegmark}

\begin{equation}
<\rho_{v}>\approx{10^{-29}}g/cm^{3}=1.63\times{10^{-27}}a.u.
\end{equation}  In other words, $<\rho_{P}>$ is too large by a factor of the order
of $10^{125}$.  This discrepancy is clearly unrealistically too large and we wish to
explore how it may be reduced. In that context, we note that to get agreement between
the theoretical value given in (4) and the observed value given in (7) requires an
$\Omega$ value as low as $2.4\times{10}^{-5}$.

First, we recall an exact calculation which we carried out to
obtain, within the framework of QED, the equation of motion of a radiating electron
\cite{ford85,ford91}.  Our analysis led to an explicit result for $\Omega$ given by

\begin{equation}
\Omega=\frac{M-m}{M\tau_{e}}, ~~~~0\le{m}\le{M},
\end{equation} where $m$ is the bare mass, $M$ is the renormalized (physical) mass,
and 

\begin{equation}
\tau_{e}=\frac{2e^{2}}{3Mc^{3}}=\frac{2}{3}~\alpha^{3}~a.u.
\end{equation}  In addition, a striking feature of our calculation was the conclusion
that causality considerations (which, in essence, state that effect follows cause 
and which lead to the technical fact that the poles of the electron response function
must lie in the lower half-plane \cite{toll,lond}) dictate that $\Omega$ has a
well-defined upper limit.  We note that the limit
$m=0$ corresponds to the largest cut-off value consistent with causality
$(\Omega=\tau_{e}^{-1})$.  Thus, \underline{causality} puts a restriction on the
maximum value of $\Omega$.  This choice had the virtue of corresponding to an
electron of minimum size and it led to a simple second-order equation for the
radiating electron, which was free of runaway solutions \cite{ford91,jack}.  However,
with respect to how large $m$ is compared to $M$, as Feynman \cite{fey}, among
others, has noted it "- - cannot be determined theoretically".  However, taking into
account retardation and relativistic effects, most estimates \cite{sak} conclude that
$m\approx(1-\alpha)M$. Substituting this result in (8) leads to the conclusion that the
cut-off, $\Omega_{QED}$ say, is given by

\begin{equation}
\Omega_{QED}\approx\alpha~\tau_{e}^{-1}=\frac{3}{2}~\alpha^{-2}=2.82\times{10}^{4},
\end{equation}  which is close to the value chosen by Bethe in his calculation of the
Lamb shift.  From (5), we see that this is smaller than $\Omega_{p}$ by a factor
$\approx{1.72}~\alpha^{21/2}=6.29\times{10}^{-23}$.  Recalling, from (4), that
$<\rho>\sim\Omega^{4}$, we see that this choice reduces $<\rho>$ by a factor of
$8.75~\alpha^{42}=1.57\times{10}^{-89}$, to now give a vaccum energy density which is
about $10^{36}$ times too large.  How can this discrepancy be reduced further? Based
on these numbers, one might expect that the energy contributed to the vacuum by the
zero-point fluctuations occurs because of the interaction between these
fluctuations and matter and not by consideration of these fluctuations in isolation. 
Clearly, more detailed work needs to be carried out but we feel that the reduction in
the discrepancy between observations and theory which stem from the above
considerations motivates more detailed analysis.

\end{document}